# Modulation of the Optical Transmittance in Multilayer Graphene by an Electrical Signal

J. L. Benítez and D. Mendoza*

Instituto de Investigaciones en Materiales, Universidad Nacional Autónoma de México. Apartado Postal 70-360, 04510 México D. F., México.

*doroteo@unam.mx

Abstract.

The modulation of the optical transmittance in multilayer graphene by means of an electrical signal in the simple configuration of coplanar electrodes is reported. Besides the fundamental frequency of modulation, higher harmonics also appear in the transmitted signal. Modulation is also observed in the optical reflectance. The modulation of the optical properties in multilayer graphene by electrical signals may be useful for the transmission of information by optical means.

## Introduction.

Carbon based materials have been intensively studied in recent decades because of their inherent physical properties. In particular, graphene is a material with a carbon atom based structure arranged in a hexagonal pattern (1). The graphene band structure allows us to make controlled changes in the Fermi level, and therefore changes in the optical transition through the gate voltage (2) (3). The unique optical properties of graphene have been studied extensively, including electro-optic modulators (4). Graphene absorbs 2.3% of visible light in a wide range of frequencies (5) and experimentally have been established that for N layers the absorption is N times that of the fundamental value (5) (6). The optical properties in the visible and far-infrared region have been studied based on the analysis of the optical conductivity using the Kubo formula (7) (8) (9) (10).

The main goal of the present paper is to report the experimental observation of the modulation of the optical transmittance in few layers graphene by an electrical signal. To knowledge, the observed phenomena has not been reported to date.

## Experimental Details.

Graphene multilayer (GM) were synthesized by the chemical vapor deposition method (CVD) (11) on copper foils (Alfa Aesar, 25 microns thickness ), using methane as the source of carbon at ambient pressure. After etching the copper foil and the washing process GM is captured on glass substrates. Two parallel and coplanar silver electrodes were placed on the GM with a length of 3mm and a separation of 5 mm.

In Figure 1(a) typical GM placed on glass substrate appears, and in Figure 1(b) the UV-Vis transmittance spectrum of a sample placed on quartz substrate is observed. Taking the value of the transmittance in the visible region we estimate 5 layer of graphene for the GM; this value was corroborated making a direct transmittance measurement with a color filter (centered at $\lambda = 620nm$) and a photodiode to detect the transmitted light (12).

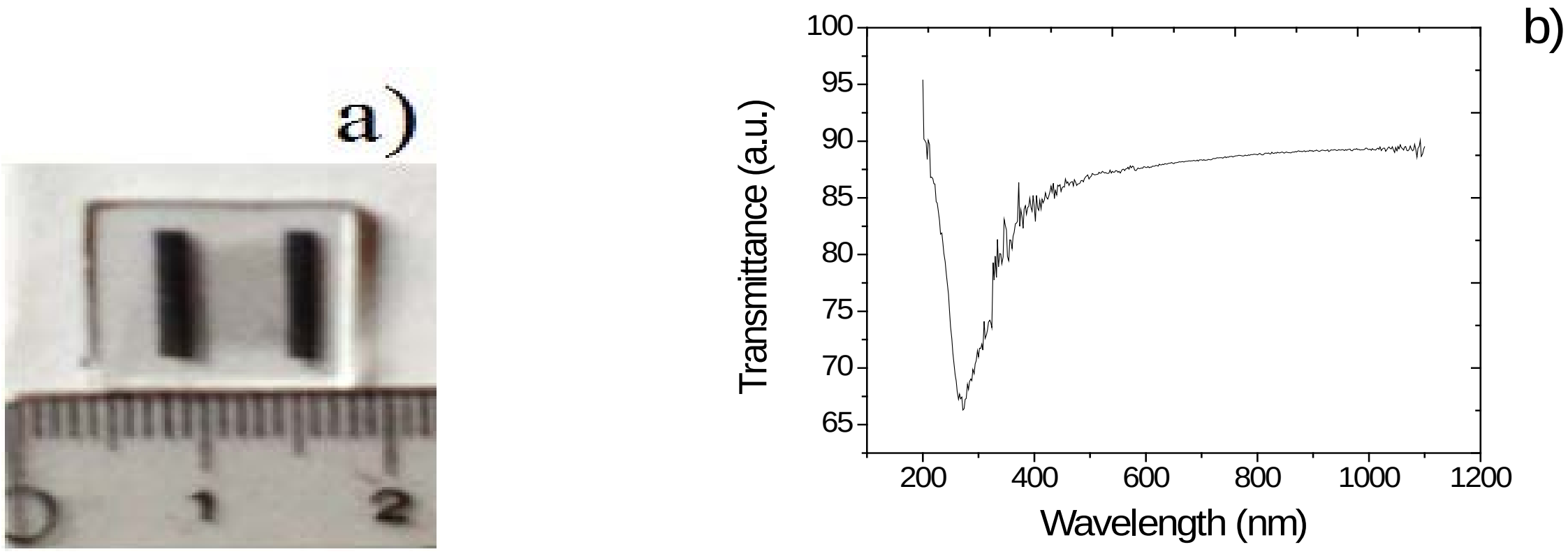

Figure 1. (a) Sample of GM with electrical contacts placed on glass substrate, (b) UV-VIS optical transmittance of similar sample placed on quartz substrate.

In Figure 2(a) a schematic diagram of the sample with electrical connections is observed and in Figure 2(b) the experimental set-up for the transmittance measurement is presented. The experiment involves passing light through the GM while the sample is biased with an electrical signal using a function generator (WAVETEK model 22) with a voltage of the form $V = V_0 sin(\omega_m t)$. The sample was illuminated with a diode laser (Edmund, diode laser 35 mW, wavelength of 785 nm, Part No. 59467), but other sources were also used. The transmitted light is detected by a photodiode (Infrared Industries, serial number 9002) and an electrometer (Keithley 619) which provides a direct physical measurement of transmittance. The reported experiments were made in vacuum ( $P = 2 \times 10^{-5} Torr$) to discard any effect by changes in the refractive index of the air due to variations in the sample temperature (see below); but many experiments were also made in air ambient.

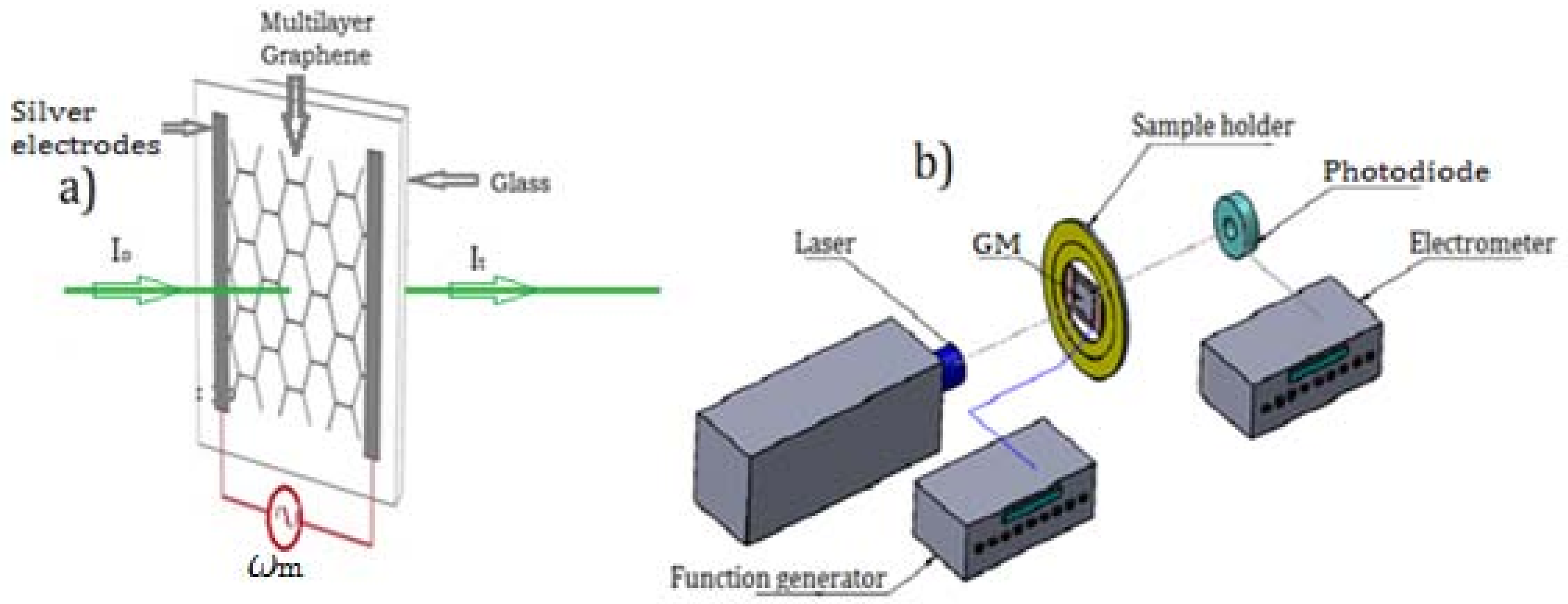

Figure 2( a) Schematic of the electrical connections and (b) illustration of the set-up for the experimental measurements.

Results and Discussion.

The overall result of the transmitted light detected by the photodiode consists of a big background signal with a small variations produced by the modulation signal, the value of the variations over the total signal varies from 1% to 5% depending on the experimental conditions. In Figure 3(a) a representative experimental result of the transmitted light is shown and in Figure 3(b) its corresponding numerical Fast Fourier Transform (FFT) in the frequency domain is presented. Note that the second harmonic signal dominates over the other harmonics and that the even harmonics appear to be more robust compared to odd harmonics. It should be noted that in some experiments only the even harmonics are visible.

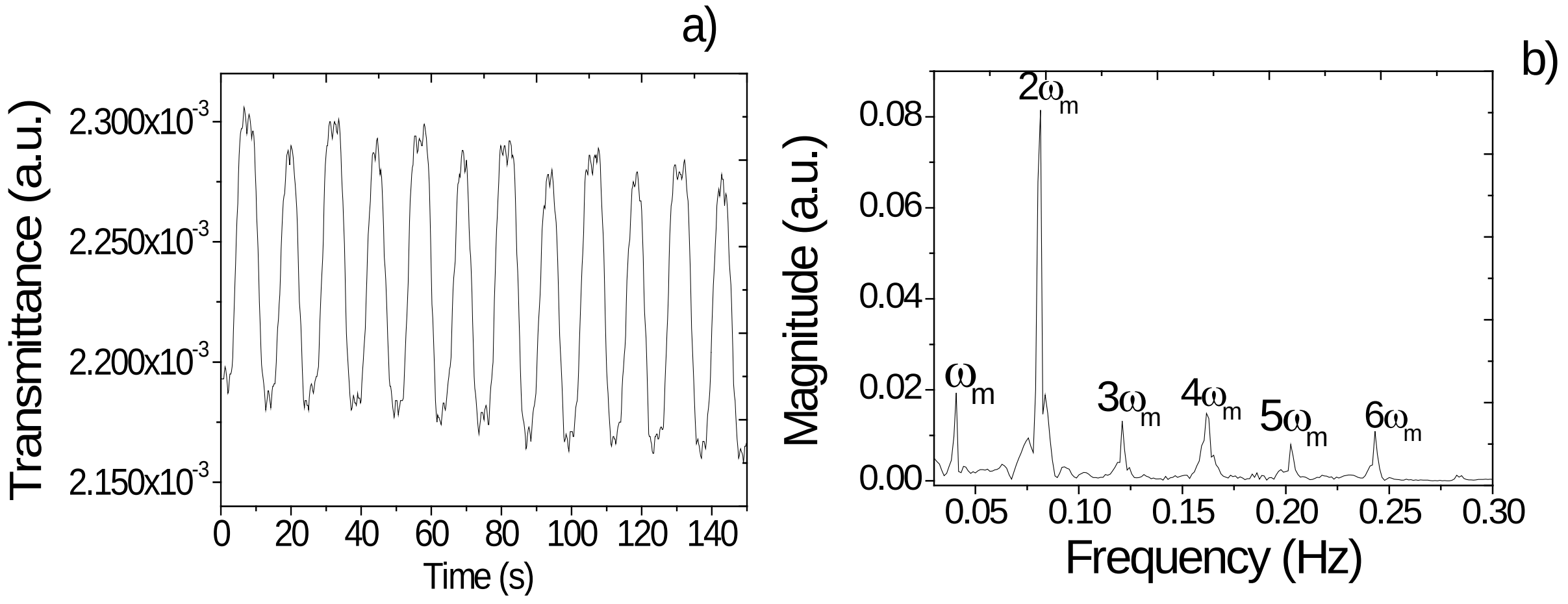

Figure 3. (a) Transmitted light intensity modulated using an amplitude of 11.2 V and a frequency of $\omega_m = 40.5\ mHz$ , (b) FFT frequency spectrum obtained from the transmittance measurements.

To explore the generality of the observed phenomena other experiments were also performed. For example, a tungsten lamp for white light was used as the source of illumination in ambient conditions. The effect of the modulation in the transmitted light is small (around 1.5%), but clearly observed as is shown in Figure 4(a), although only the first and the second harmonics are present in the FFT ( Figure 4(b)). Note again that the second harmonic signal is greater than the first one.

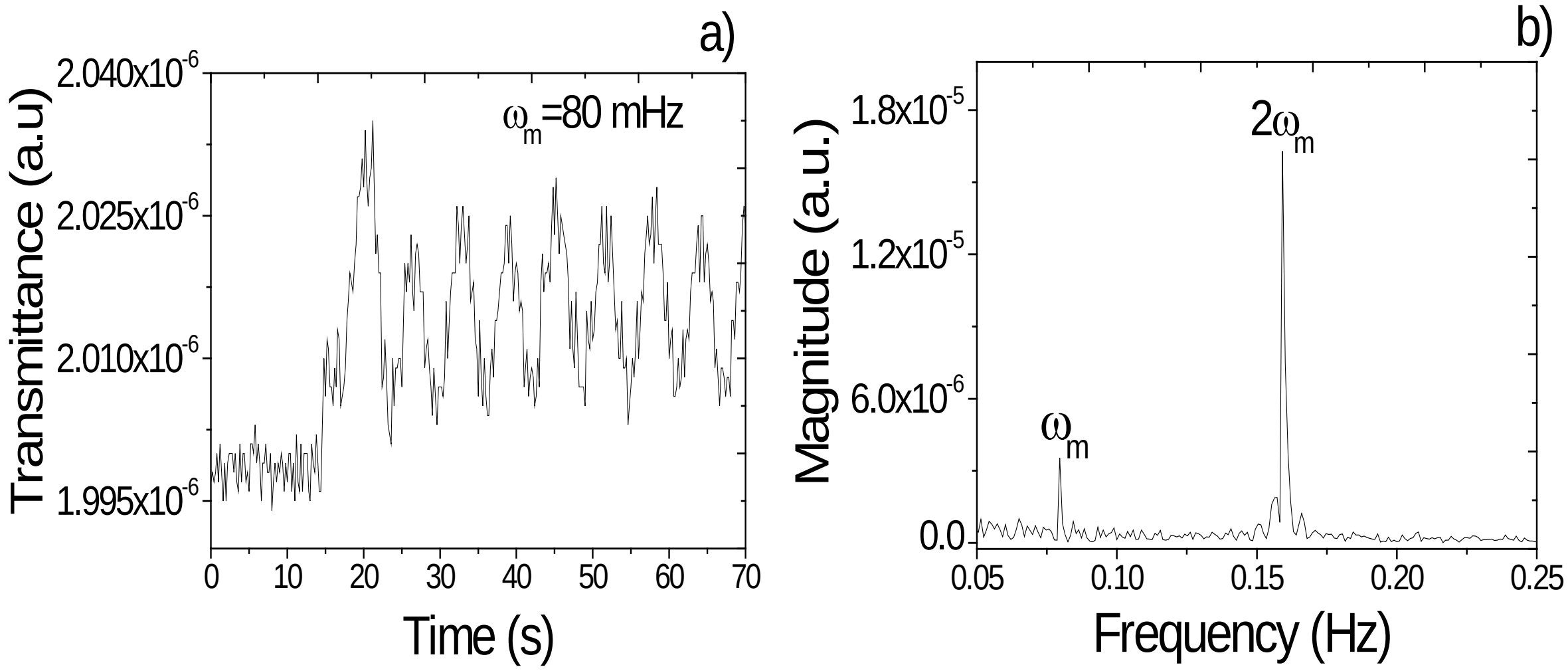

Figure 4. (a) Transmitted intensity using white light with a modulation frequency of 80 mHz and an amplitude of 11.2V, (b) FFT spectra of the transmitted signal.

The other two experiments are: change in the transmittance when the sample is biased with a constant voltage (DC) as is shown in Figure 5(a) and modulation in the reflected light as is observed in the numerical FFT of the signal presented in Figure 5(b). Both experiments were made in ambient conditions.

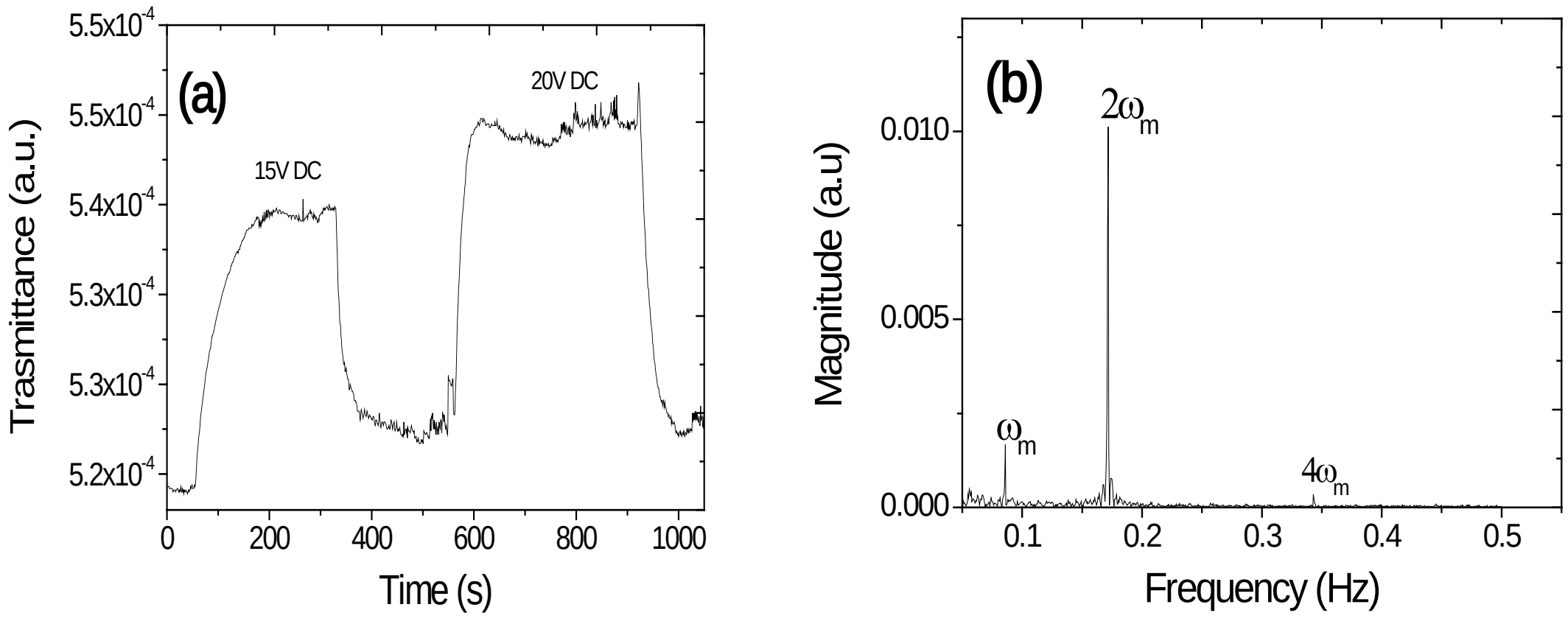

Figure 5. Changes in the transmittance when the sample is biased with a DC signal, (b) FFT of the reflected light using λ=785 for the illumination and a modulation frequency of $\omega_m = 85\ mHz$.

In order to get some insight on the physical mechanism involved in the observed phenomena, we explore two possible scenarios: one in which the electrical permittivity in graphene varies as a function of the electrical field (model 1) and the second where the optical conductivity depends on the temperature of the sample (model 2).

Model 1. Santos and Kaxiras by means of first-principles electronic calculations have shown that the static electrical permittivity can be tuned by an external electric field due to the charge polarization (13).They find that both the out-of-plane and in-plane components of the permittivity change in multilayer and one layer ribbons, respectively. Although the calculations were made for one layer graphene in the in-plane case, we generalize those results to the experimental situation where few layer graphene with coplanar electrical contacts were used (recall: the electrical modulation signal is applied in the plane of the sample).

The optical transmittance Tr is calculated using the expression $Tr = exp(-\alpha d)$ which has been used for multilayer graphene of thickness d, where α is the absorption coefficient given by $\alpha = \frac{4\pi}{\lambda} n_i$ , $n_i$ being the imaginary part of the refractive index and λ the wavelength of the transmitted light (14). The last parameter can be expressed in terms of the real ( $\varepsilon_r$ ) and imaginary ( $\varepsilon_i$ ) parts of the electric permittivity as follows:

$$n_i = \left(\frac{1}{2}\left(-\varepsilon_r + \left(\varepsilon_r^2 + \varepsilon_i^2\right)^2\right)\right)^{1/2} \qquad (1)$$

To apply the results of reference (13), we suppose that the dependence on the electric field of the electrical permittivity is only through $\varepsilon_r$ and that $\varepsilon_i$ is a constant. As a first approximation we suppose a linear dependence on the electric field E in units of V/Å, and as a rough estimation we take the points (E, $\varepsilon_r$ )=(0.1,11), (0.5, 15) taken from the calculated data of the Figure 1(c) of reference (13) for the wider ribbon; these give a straight line of the form:

$$\varepsilon_r = 10 + 10E = 10 + |10\, E_0 sin(\omega_m t)|, \qquad (2)$$

where in the last equality $E = E_0 sin(\omega_m t)$ is the in-plane modulation electric field which is taken in its absolute value because of the symmetry in both electrodes in the coplanar configuration. For a typical experimental condition, $E_0$=11 V/5 mm= 2.2x10$^{-7}$ V/Å, d=1.7nm corresponding to 5 layers of graphene, and $\omega_m = 40.5mHz$ which is in the static limit compared to the optical frequencies used for the transmitted light with wavelength λ. Taken the value $\varepsilon_i$ = 12 (15) and λ=785 nm, and substituting eq. (2) in to eq. (1), in Figure 6 the calculated optical transmittance (a) and its corresponding numerical Fourier transform (b) are presented. Although the absolute value of the calculated transmittance does not correspond to the expected value for 5 layers, the appearance of even harmonics of the fundamental modulation frequency $\omega_m = 40.5mHz$ is clear from the FFT result.

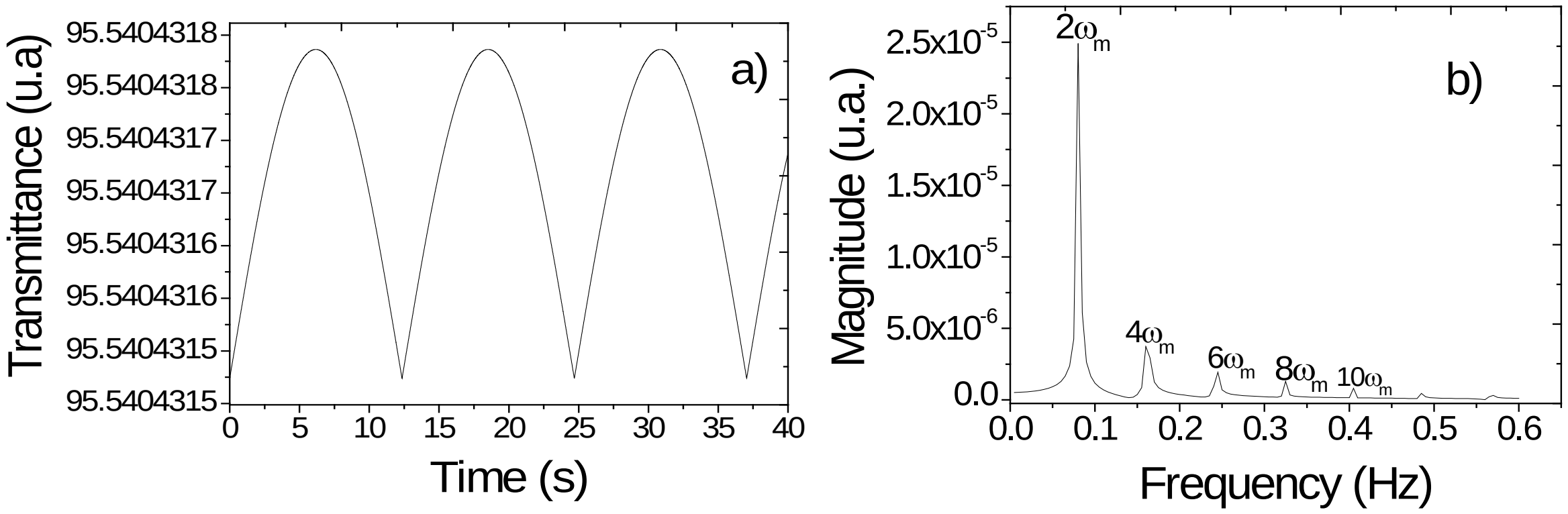

Figure 6. (a) Transmittance obtained using d = 1.7 nm, $\omega_m = 40.5\ mHz$ and λ = 785 nm; (b) numerical Fourier transform of the transmittance given in (a).

Model 2. The second scenario is based on the fact that the sample heats via the Joule effect when it is biased with the modulation signal. The temperature of the sample is modulated in time at twice the frequency of the modulation signal and in the proper conditions the thermo acoustic effect in our samples is observed (16). In this model we follow the approach of the temperature (T) dependence of the optical conductivity. In the visible region of the spectrum, where the transmittance experiments were made, the inter-band contribution is the most important part of the optical conductivity (17) and is given by:

$$Re\sigma^{inter}(\omega) = \frac{e^2}{4\hbar}\frac{\sinh\left(\frac{\hbar\omega}{2k_BT}\right)}{\cosh\left(\frac{E_F}{k_BT}\right)+\cosh\left(\frac{\hbar\omega}{2k_BT}\right)}, \tag{3}$$

where e is the electron charge, $k_B$ the Boltzmann constant, $\hbar\omega$ is the photon energy, and $E_F$ the Fermi energy whose temperature dependence for graphene is given by $E_F = E_{F0} - \frac{\pi^2}{6}\frac{(k_BT)^2}{E_{F0}}$.

For the case of multilayer graphene, the optical transmittance can be expressed as (18):

$$Tr = \frac{1}{\left(1+\frac{NZ\sigma(\omega)}{1+n}\right)^2}, \tag{4}$$

where N denotes the number of graphene layers, $Z = 377\Omega$ is the free-space impedance, n=1.52 is the refractive index of the substrate, and $\sigma(\omega)$ corresponds to the optical conductivity of one layer of graphene (18). In the case of Joule heating, the temperature is proportional to the electrical power dissipated in the sample, if the modulation voltage goes as $V = V_0 sin(\omega_m t)$, then the temperature can be written as

$$T(K) = T_0 + T_m sin^2(\omega_m t) = 310 + 40 sin^2(\omega_m t), \tag{5}$$

where in the last equality of eq. (5) the numerical parameters were chosen according to a typical experimental condition. The calculation of the optical transmittance, as well as its corresponding numerical Fourier transform for 5 layer of graphene are presented in Figure 7.

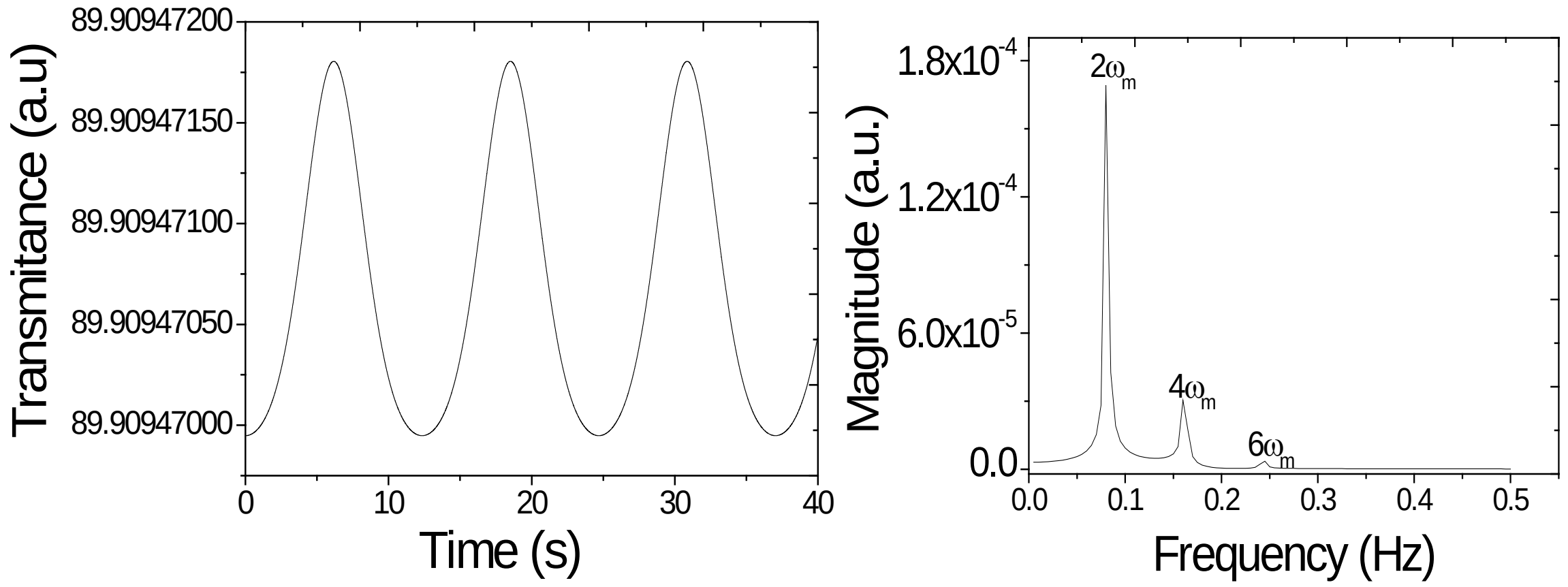

Figure 7. (a) Transmittances against time obtained for $E_{F0} = .5\ eV$, $\omega_m = 40.5\ mHz$ and λ = 785 nm; (b) numerical Fourier transform of the transmittance given in (a).

Note that in both models only even harmonics of the fundamental modulation frequency are reproduced. We believe that both scenarios may be possible in the experimental conditions of the observed phenomena, but the origin of the odd harmonics is not clear to date. One possibility may be related with the existence of non linear response of electrical carriers under the presence of an electromagnetic field (19), (20). For example, Mikhailov (19) studied the non linear electrodynamic response of carriers in graphene using the Boltzmann transport equation. His calculations, valid for intraband transitions and low frequencies, lead to the existence of electrical currents with only odd harmonics of the excitation frequency. The apparition of these currents with odd frequency may be related with the odd harmonics observed in our experiments, but more studies are necessary to clarify this issue.

Conclusion.

The optical transmittance of few layers of graphene can be modulated by means of an electrical signal in a very simple configuration of two coplanar electrical contacts. Apart of the signal with the fundamental modulation frequency, higher harmonics are also observed. The observed effect appears to be a universal phenomena because it works with monochromatic and white light, with a DC bias, and in the reflection mode as well. As a first approximation to explain the phenomena, we explore two possibilities: one in which the electrical permittivity depends on the electric field and the second where the optical conductivity is affected by the temperature. Using both models the optical transmittance were calculated and the existence of even harmonics is shown. More theoretical work is necessary in order to explain the existence of odd harmonics. Although our experiments were made in the range of mHz for the modulation frequency, we believe that the range can be extended to higher frequencies. The observed phenomena opens the possibility of using this effect in the transmission of information by optical means and for the generation of electrical signals with higher frequencies than that of the excitation signal.

Acknowledgement. We thank R. Castañeda of CCADET-UNAM for providing us with the diode laser.